\documentclass[3p,twocolumn,preprint]{elsarticle}

\usepackage{hyperref}
\usepackage{multirow}
\usepackage{textcomp}
\usepackage{amsmath}

\journal{npj 2D Materials and Applications (DOI: 10.1038/s41699-019-0088-4)}

\biboptions{sort&compress}
\begin{document}

\begin{frontmatter}

\title{Mechanism of substrate-induced anisotropic growth of monolayer WS$_2$ by kinetic Monte Carlo simulations}

\author{Lixiang Wu}
\ead{wulx@hdu.edu.cn}

\author{Weihuang Yang}
\ead{yangwh@hdu.edu.cn}

\author{Gaofeng Wang}
\ead{gaofeng@hdu.edu.cn}

\address{School of Electronics and Information, Hangzhou Dianzi University, Hangzhou 310018, China}

\begin{abstract}
Controlled anisotropic growth of two-dimensional materials provides an approach for the synthesis of large single crystals and nanoribbons, which are promising for applications as low-dimensional semiconductors and in next-generation optoelectronic devices. In particular, the anisotropic growth of transition metal dichalcogenides induced by the substrate is of great interest due to its operability. To date, however, their substrate-induced anisotropic growth is typically driven by the optimization of experimental parameters without uncovering the fundamental mechanism. Here, the anisotropic growth of monolayer tungsten disulfide on an ST-X quartz substrate is achieved by chemical vapor deposition, and the mechanism of substrate-induced anisotropic growth is examined by kinetic Monte Carlo simulations. These results show that, besides the variation of substrate adsorption, the chalcogen to metal (C/M) ratio is a major contributor to the large growth anisotropy and the polarization of undergrowth and overgrowth; either perfect isotropy or high anisotropy can be expected when the C/M ratio equals 2.0 by properly controlling the linear relationship between gas flux and temperature.
\end{abstract}


\end{frontmatter}


\section*{INTRODUCTION}
\noindent Transition metal dichalcogenides (TMDCs) have been a star family of two-dimensional (2D) materials~\cite{manzeli20172d}, which is attributed to their excellent elctronic and optical properties.
In particular, monolayer TMDCs typically including MoS$_2$, MoSe$_2$, WS$_2$, and WSe$_2$, with remarkable advantages over the few-layers or blocks of such materials, have more potential applications in low-dimensional semiconductors and next-generation optoelectronic devices.~\cite{kumar2018scalable,lu2017janus,hsu2017topological,christiansen2017phonon,steinhoff2017exciton,duan2015two}
Methods including mechanical exfoliation, liquid exfoliation, physical vapor deposition, and chemical vapor deposition (CVD) have been developed to prepare monolayer or few-layers TMDC materials.~\cite{choi2017recent,wong2016recent}
Among those methods, vapor-phase-based growth approaches like CVD are more desirable due to the potential to scale up and obtain wafer-scale 2D TMDCs.

To enable the direct growth of large single crystals and nanoribbons by CVD~\cite{zhang2018reliable}, the flake alignment of 2D materials can be one of key factors and the growth anisotropy should be intensively investigated, since it has been shown that the wafer-scale single-crystalline graphene can be grown if the initial graphene nuclei have the same orientation~\cite{tsivion2011guided} and the alignment of monolayer TMDC nanoribbons is largely determined by the orientation of the crystal substrate.~\cite{chen2017fabrication,li2018vapour}
Currently~\cite{artyukhov2016topochemistry}, in many synthesis approaches, 2D TMDC materials nucleate randomly on substrates, and their orientation cannot be well controlled.
Controlled anisotropic growth of large-area or high-aspect-ratio single crystals is still in the early stage of development, and the detailed mechanism remains unclear, though there have been efforts and attempts on the location- and orientation-controlled growth of monolayer TMDCs.~\cite{zhang2013controlled,shi2015recent,bersch2017selective}

Recently, substrates such as sapphire, mica, graphite~\cite{chen2015step,qin2017van,hwang2017te,bonilla2018strong} have been used in the alignment of as-grown TMDC flakes, which is attributed to the van der Waals (vdW) epitaxial interaction between the crystal substrate and monolayer TMDCs.
The substrate-induced growth anisotropy of 2D TMDC flakes has been found in the experiments.
For example, Chen \textit{et al.} reported a step-edge-guided approach for the aligned or oriented growth of 2D WSe$_2$ on the C-plane sapphire substrate by CVD and found that at a high temperature ($>$950 \textdegree C), the growth is strongly guided by the atomic steps on the substrate surface;~\cite{chen2015step} using graphene on Ir(111) as substrates, Hall \textit{et al.} grew well-oriented monolayer flakes of TMDCs by a two-step molecular beam epitaxy synthesis.~\cite{hall2018molecular}
However, the substrate-induced anisotropic growth of 2D TMDC samples is almost driven by the optimization of experimental parameters without understanding the fundamental mechanism dictating the 2D domain morphology under diverse growth conditions.

Systematic understanding of the anisotropic growth of monolayer TMDCs is a theoritical challenge due to the diversity of involved kinetic mechanisms and the wide range of growth conditions (\textit{e.g.}, gas flux, C/M ratio, temperature, and substrate conditions).
The kMC simulation~\cite{battaile2002kinetic,hoffmann2014kmos}, as an excellent tool to investigate the cumulative statistical effects of the kinetic processes at the atomic level, can help quantify these diverse experimental conditions and significantly reduce the number of variables, so that one can develop a unified conceptual framework on the deposition mechanism of 2D compound crystals.
Rajan \textit{et al.} proposed a generalized kMC model with special consideration to CVD reactor parameters for the growth of 2D TMDC monolayers, which is predictive of mophological evolution with variations of growth conditions~\cite{govind2016generalized}.
Nie \textit{et al.} introduced a full-diffusion kinetic Monte Carlo (kMC) model coupled with first-principles calculations to study the deposition process of WSe$_2$ monolayers on graphene,~\cite{nie2016first} which can reproduce different morphologies such as compact, fractal, and dendrite.
Nevertheless, substrate effects~\cite{sun2017substrate,ahn2017strain} that can largely determine the growth anisotropy are not introduced to those kMC models.

Study on the substrate-induced anisotropic growth of monolayer WS$_2$ can provide a complete perspective closer to reality on the growth of 2D materials.
In this work, by introducing local substrate effects on the adsorption, desorption, and diffusion processes, a substrate-sensitive kMC model is established to study the growth anisotropy using the representative case of the CVD growth of monolayer WS$_2$ on ST-X quartz.
And the kMC model introduces the on-the-fly calculation of substrate-induced variation of atomic potential energy.
Then the growth morphological anisotropy and its origin are analyzed and discussed with experimental observations and simulation results.

\section*{RESULTS AND DISCUSSION}

\subsection*{Observation and measurement of growth anisotropy}

\begin{figure}[tbp]
  \centering
  \includegraphics[width=\columnwidth]{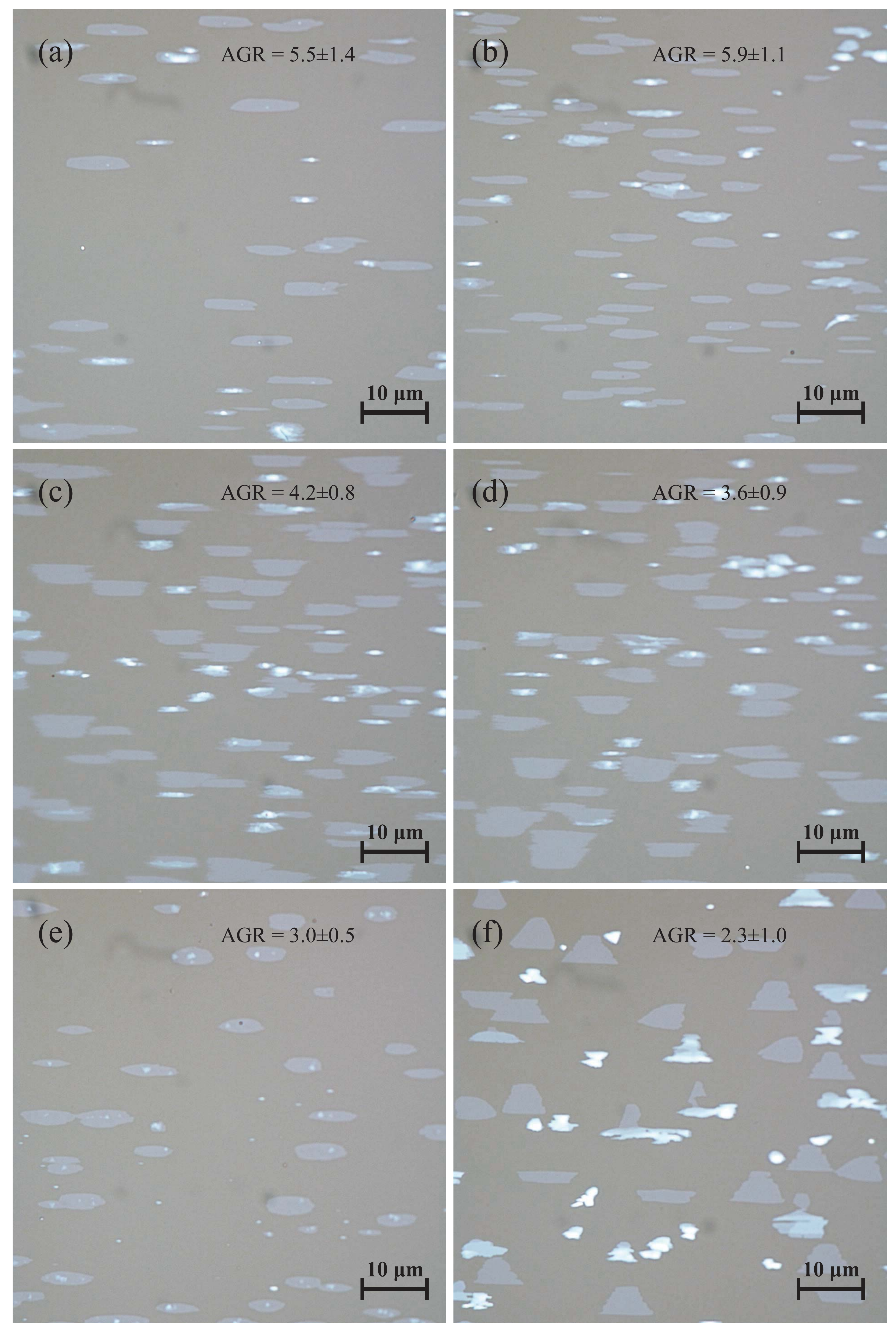}
  \caption{growth of monolayer WS$_2$ on the ST-X quartz substrate; (a) and (b) are two neighbor spots at the front side of sample \#1; (c) and (d) are two neighbor spots at the rear side of sample \#1; (e) and (f) are two separated spots at the rear side of sample \#2.}
  \label{fig:GrowthOfWS2}
\end{figure}

\noindent We conducted a preliminary experiment and observed the anisotropic growth of monolayer WS$_2$ before the experimental optimization.
It has been found that the morphology of TMDC flakes varies from upward triangle to downward triangle and to hexagon due to the variation of conditions such as temperature (T), gas flux (Ra, defined as the adsorption rate of W atoms in the kMC model), the C/M ratio, and substrate adsorption contrast (xEads, the ratio of the highest adsorption energy to the lowest on the same substrate).
The experimental is detailed in the section of METHODS.
Figure~\ref{fig:GrowthOfWS2} shows the growth of monolayer WS$_2$ on the ST-X quartz substrate at specific locations, where (a) and (b) are photographed with amplification of 1000 at two upstream neighbor spots of sample \#1; (c) and (d) are photographed with amplification of 1000 at two downstream neighbor spots of sample \#1; (e) and (f) are photographed with amplification of 1000 at two separated spots downstream from sample \#2.
Specifically, sample \#1 is placed at the downstream of the S source and sample \#2 is arranged at the further downstream, which is 1.5~cm away from sample \#1.
Figure~\ref{fig:GrowthOfWS2} shows a high degree of consistency in orientation of WS$_2$ flakes and, particularly, Figure~\ref{fig:GrowthOfWS2}f shows a morphological evolution of WS$_2$ flakes from trapezoid to triangle.

\begin{figure}[t]
  \centering
  \includegraphics[width=0.75\columnwidth]{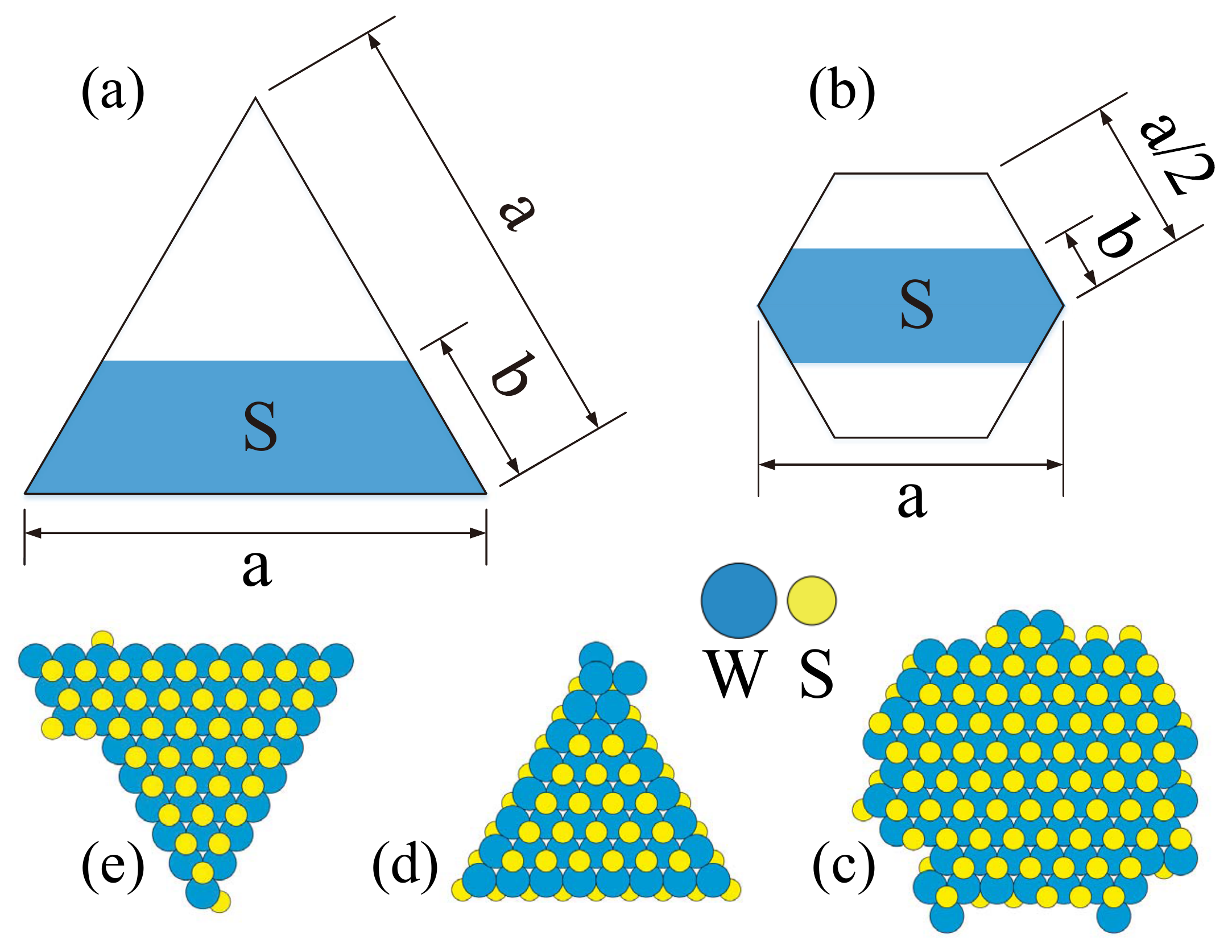}
  \caption{(a) and (b) illustrate how to calculate the AGRs of monolayer WS$_2$ flakes with different shapes, including (c) hexagon, (d) upward triangle, and (e) downward triangle, where the downward trianglar flake is terminated by 3 W-zigzag (-zz) edges, the upward by 3 S-zz edges, and the hexgonal by both 3 W-zz and 3 S-zz edges.}
  \label{fig:FlakeMorph}
\end{figure}

Quantitative analysis is applied to observation and measurement of the anisotropic growth of 2D TMDC flakes.
To make the growth anisotropy measurable, anisotropic growth ratio (AGR) is defined to describe the extent of growth anisotropy and the formula for calculating AGRs is as follows:
\begin{align}
  P(a, S) = \begin{cases}
    \frac{1}{1 - \sqrt{1 - \frac{4 S}{\sqrt{3} a^2}}} & \text{for triangle} \\
    \frac{1}{2} \frac{1}{1 - \sqrt{1 - \frac{2 S}{\sqrt{3} a^2}}} & \text{for hexagon}
  \end{cases},
  \label{eqn:AnisoDef}
\end{align}
where $a$, $b$, and $S$ are the polygon diameter, grown length along a side, and area of the TMDC flake, respectively.
Figure~\ref{fig:FlakeMorph}a/b illustrate the AGR calculation of a triangular or hexagonal TMDC flake by measuring the three parameters in (\ref{eqn:AnisoDef}).
Figure~\ref{fig:FlakeMorph}c-e show three representative morphologies of monolayer WS$_2$ flakes having AGRs of about 1.0, which are grown with uniform substrate adsorption under different growth conditions.

The morphological classification of TMDC flakes is conducted to extract the growth tendency and make the high-dimensional simulation data readable, where the flake with the area ratio (to the defined substrate lattice) greater than 20\% is set as overgrowth while the one with the area ratio less than 3\% as undergrowth according to rules of thumb.
Similarly, the flake with AGR~$>$~4.0 is considered as (extremely) anisotropic growth whereas the flake with AGR around 1.0 as isotropic growth.
For example, Figure 1a-c show that the average AGRs of monolayer WS$_2$ flakes are greater than 4.0, indicating the extremely anisotropic growth occured.

\subsection*{Substrate-sensitive kMC model}
\begin{figure}[t]
  \centering
  \includegraphics[width=\columnwidth]{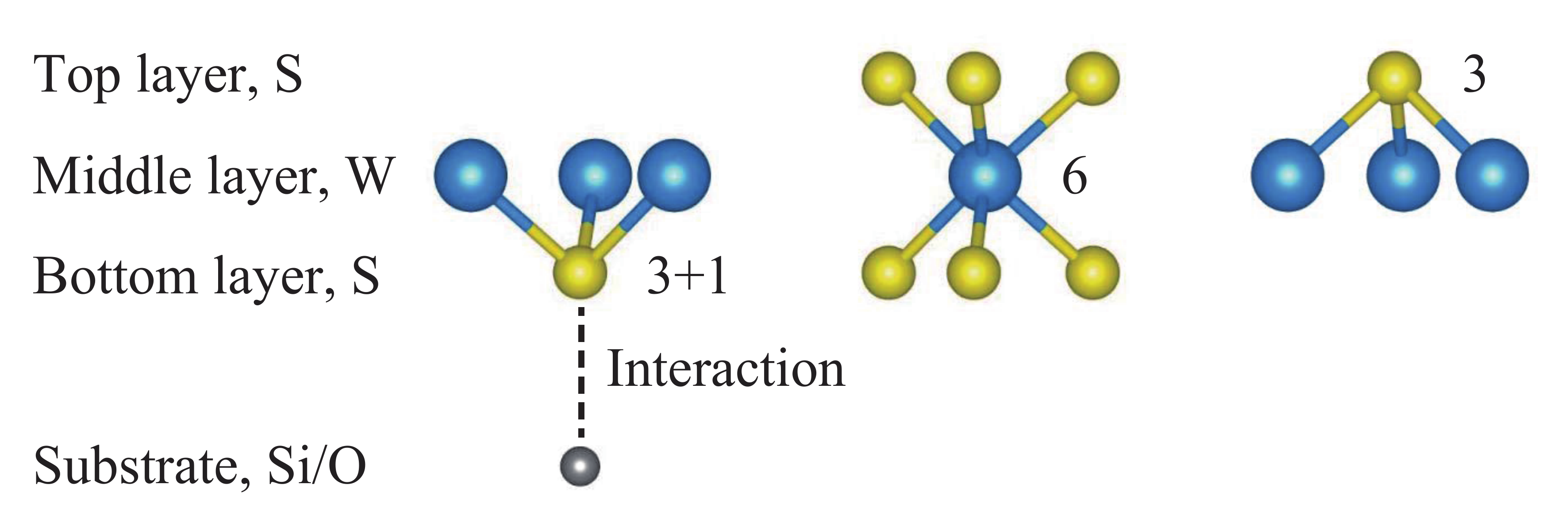}
  \caption{the atomic arrangement and bonding environment with first order approximation.}
  \label{fig:DiffModel}
\end{figure}

\noindent To study the anisotropic growth of 2D TMDCs, we propose a full-diffusion kMC model coupling with local substrate effects including substrate adsorption and surface diffusion, which are considered as two key factors that influence the topological evolution of TMDC monolayers and would result in the anisotropy of film growth.
Compared with the bond-counting model of first order approximation, as shown in Figure~\ref{fig:DiffModel}, our model additionally does the on-the-fly counting of interactions between source atoms (W or S) and substrate atoms (Si or O).
Therefore, substrate-induced variations of adsorption energy and diffusion energy are calculated in real-time and then, specifically, the model can simulate the anisotropic growth of WS$_2$ on the ST-X quartz substrate.

Before the kMC simulations, we conducted first-principles calculations to obtain energy parameters for modeling the above-mentioned kinetic events.
Without loss of generality, the growth of WS$_2$ monolayers on a ST-X quartz substrate is investigated.
The calculated results are presented in Table~\ref{tbl:EnergyParameters}.
\begin{table}[tb]
\small
  \caption{\ Energy parameters (unit: eV)}
  \label{tbl:EnergyParameters}
  \begin{tabular*}{0.48\textwidth}{@{\extracolsep{\fill}}lcc}
    \hline
    \multirow{3}{*}{Bond energy} & W--W & 1.16 \\
     & W--S & 2.14 \\
     & S--S & 0.329 \\
    \hline
    \multirow{4}{*}{Diffusion energy} & W (edge) & 3.81 \\
     & S (edge) & 1.02 \\
     & W or S (surface) & 0.03 -- 0.05 \\
     & W or S (interlayer) & 0.02 \\
     \hline
     \multirow{2}{*}{Adsorption energy} & W (substrate) & -2.0 -- -1.0\textsuperscript{\emph{$\ast$}} \\
      & S (substrate) & -1.28 -- -0.64\textsuperscript{\emph{$\ast$}} \\
     \hline
  \end{tabular*}

  \textsuperscript{\emph{$\ast$}} The adsorption energy varies periodically in a specific range in corresponding to the periodical sawtooth profile of the ST-X quartz surface (see Figure~\ref{fig:ModelDesc}).
\end{table}

\begin{figure}[htb]
  \centering
  \includegraphics[width=0.75\columnwidth]{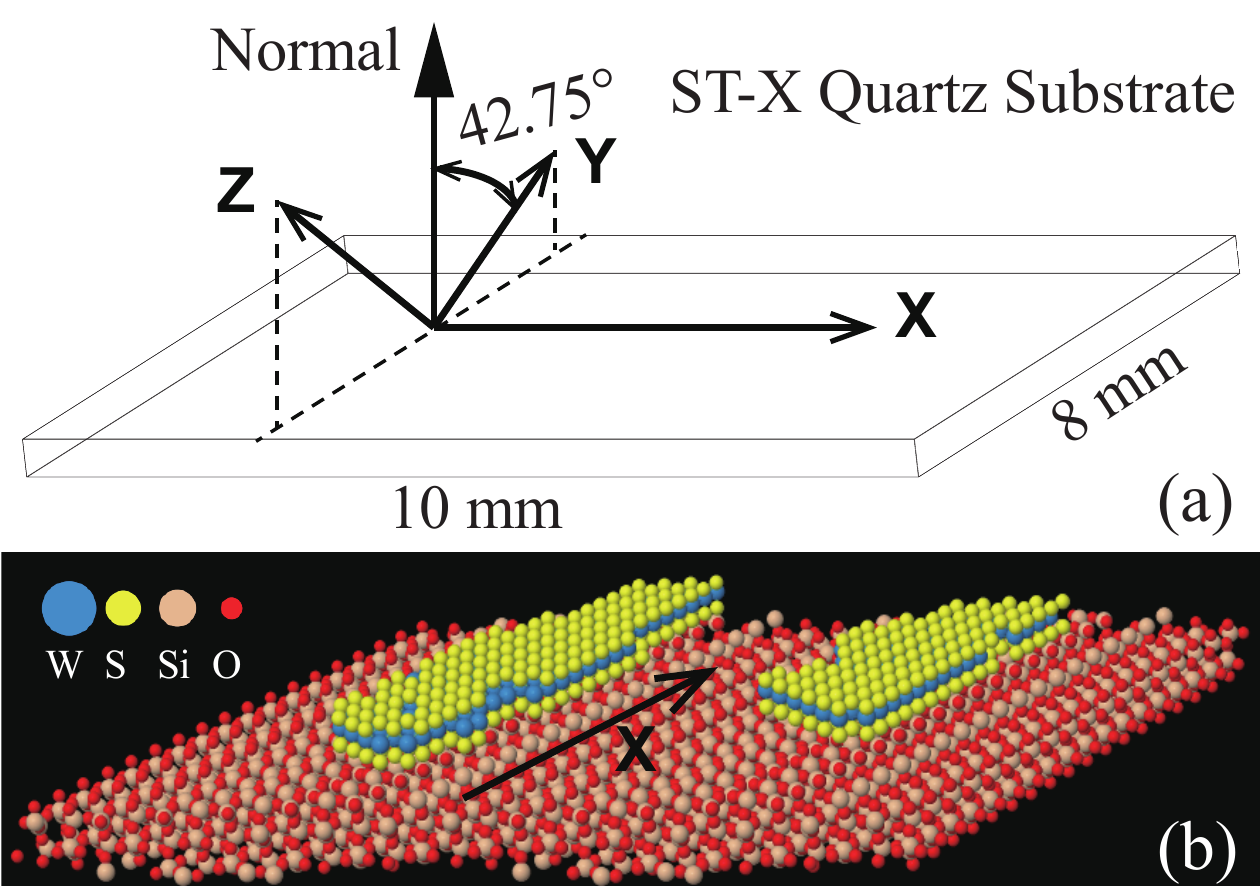}
  \caption{(a) the cut specification of ST-X quartz substrate and (b) the orientation of as-grown WS$_2$ flakes with respect to the substrate.}
  \label{fig:SubOrient}
\end{figure}

\begin{figure}[t]
  \centering
  \includegraphics[width=\columnwidth]{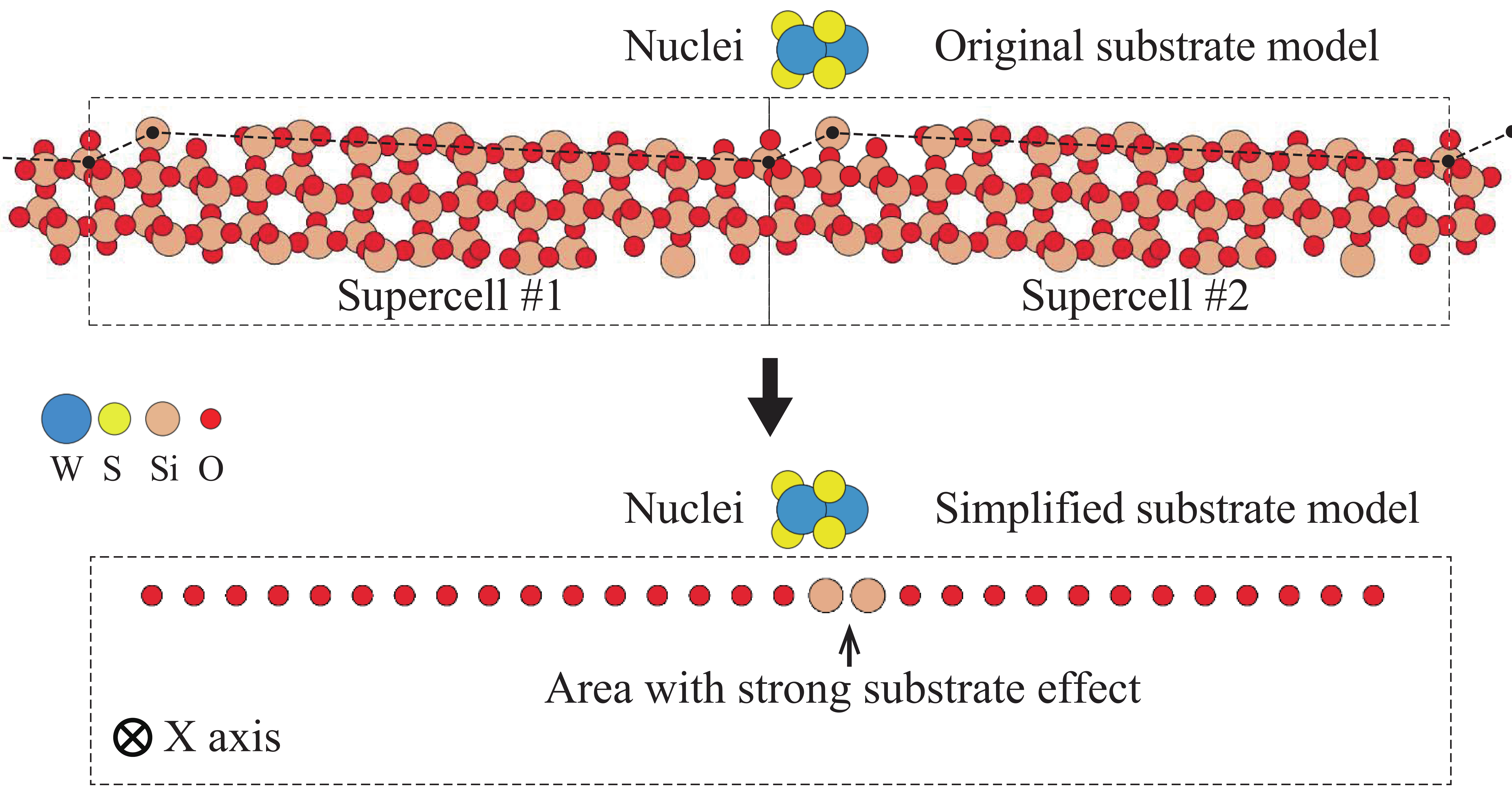}
  \caption{the original atomic structure of WS$_2$ on the ST-X quartz substrate and its simplified configuration.}
  \label{fig:ModelDesc}
\end{figure}

ST-X quartz, also called 42.75\textdegree~Y-X quartz (see Figure~\ref{fig:SubOrient}a), has been widely used as the substrate of surface acoustic wave (SAW) resonators.~\cite{parker1988precision}
Because the normal to ST-X quartz surface is 42.75\textdegree~rotated with respect to axis Y, there is a periodical sawtooth profile with a slightly tilt on the ST-X quartz surface (see Figure~\ref{fig:ModelDesc}) and the vdW interaction between the WS$_2$ and substrate varies peridically along the sawtooth profile.
The orientation of as-grown WS$_2$ flakes in Figure~\ref{fig:GrowthOfWS2} is consistent with the X axis of ST-X quartz substrate, which is illustrated in Figure~\ref{fig:SubOrient}b.
It should be mentioned that the temperature in the CVD process of monolayer WS$_2$ is much higher than the phase-transition temperature of quartz crystals (around 550\textdegree C), so the ST-X quartz is in $\beta$ phase during the growth process.

For simplicity, as illustrated in Figure~\ref{fig:ModelDesc}, the ST-X quartz substrate is reduced into a binary model that only has two types of substrate adsorption domain: strong adsorption domain marked by Si and ordinary adsorption domain marked by O.
With the reduced configuration of the ST-X quartz substrate, it can simplify the kMC computations without loss of physical content.
The seed for the initialization of the kMC simulation is taken to be a hexagonal W$_3$S$_6$ nucleus.
The simplified substrate model is applied to the following kMC simulations, which is in default a 30$\times$30 lattice with a 2$\times$30 `belt' marked by Si of strong substrate adsorption compared to other areas marked by O of ordinary substrate adsorption.

\subsection*{Simulation of substrate-induced anisotropic growth}
\noindent Three groups of kMC simulations are designed and presented in Table~\ref{tbl:SimConds}.
The anisotropic growth and isotropic growth co-occur when the substrate adsorption is neither large nor small (xEads~$=$~1.25 or 1.5), which agrees with the observations in the preliminary experiment, so the substrate adsorption contrast is estimated to be 1.4.

\begin{table}[h]
\small
  \caption{\ Growth conditions for kMC simulations}
  \label{tbl:SimConds}
  \begin{tabular*}{0.48\textwidth}{@{\extracolsep{\fill}}ccccc}
    \hline
    No. & T (K) & Ra (s$^{-1}$) & C/M ratio & xEads \\
    \hline
    1 & 973$\to$1273 & 9 & 0.5$\to$4 & 1.4 \\
    2 & 1123 & 0.6$\to$18 & 0.5$\to$4 & 1.4 \\
    3 & 973$\to$1273 & 0.6$\to$18 & 2 & 1.4 \\
    \hline
  \end{tabular*}
\end{table}

\begin{figure*}[t]
  \centering
  \includegraphics[width=0.75\textwidth]{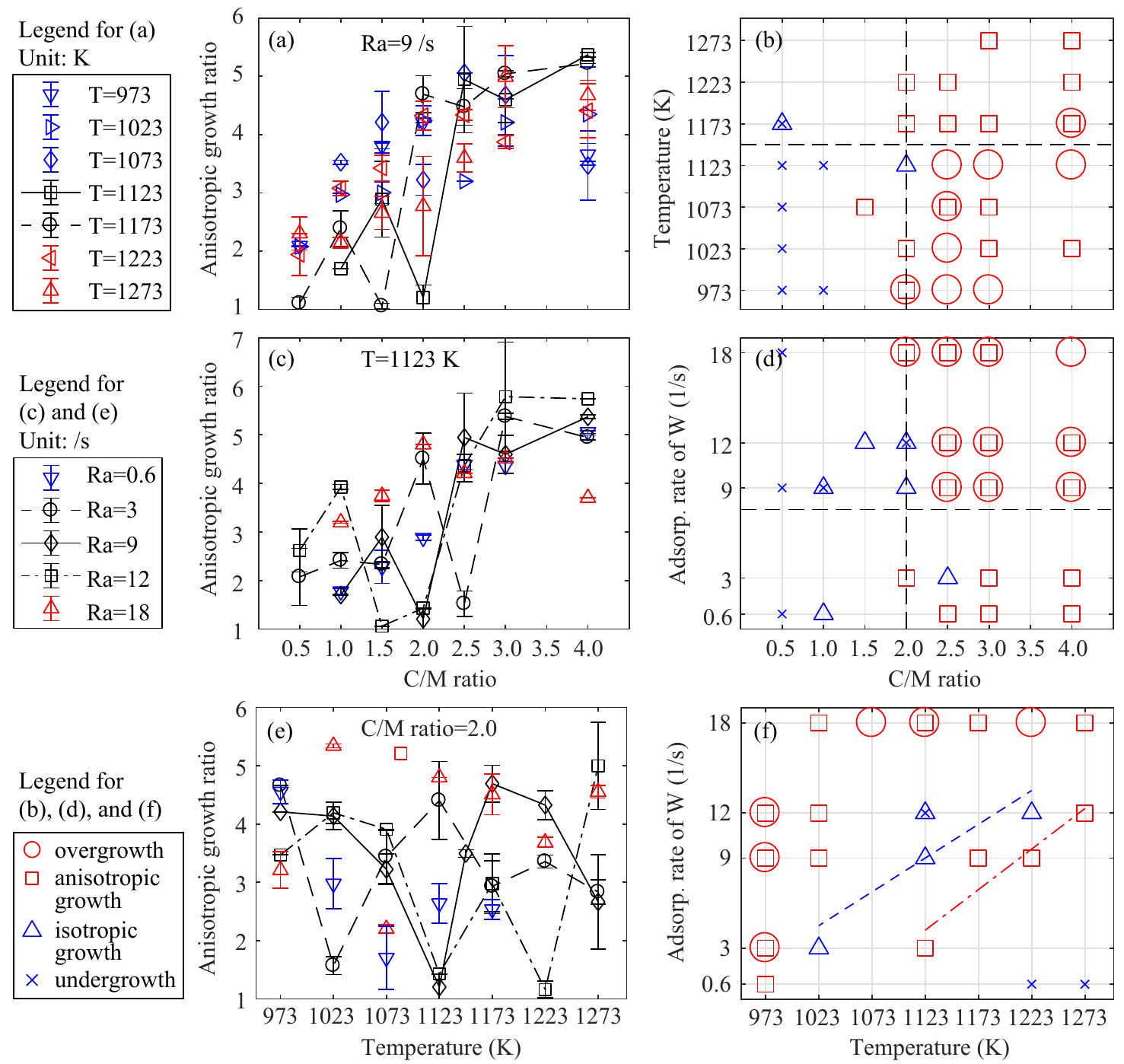}
  \caption{results of the kMC simulations Nos 1, 2, and 3. (a), (c), and (e) are AGR measurement curves; (b), (d), and (f) are morphological classification maps.}
  \label{fig:MorphVar}
\end{figure*}

As is shown in Figure~\ref{fig:MorphVar}, the C/M ratio is possibly a major contributor to the surge of growth anisotropy and the polarization of undergrowth and overgrowth; the specific linear relationship between gas flux and temperature can help growth isotrophy or anisotropy maintain in an expected manner.
The AGR surges at the vicinity of the C/M ratio of 2.0, as is shown in Figure~\ref{fig:MorphVar}a/c, and too high or too low temperature (or too small or too large gas flux) can weaken the dramatic increase, which is observed at the AGR curves of T~$=$~1123, 1173 and Ra~$=$~3, 9, 12.
Figure~\ref{fig:MorphVar}b/d show that anisotropic growth is observed as the C/M ratio increases and reaches over 2.0, while at the same time very low temperature or very large gas flux would result in overgrowth.
At the dashed line of the C/M ratio of 2.0, the morphological classification is diverse and the morphology varies with the change of either temperature or gas flux.
Moreover, Figure~\ref{fig:MorphVar}c/e provide more details for the calculated results when the C/M ratio equals to 2.0.
It is clearly shown that WS$_2$ monolayers grow isotropically only in the center area where both the temperature and gas flux are not too low or too high, and the anisotropic growth of WS$_2$ monolayers also has a similar ``safe zone''.
After linear regression calculations, it is found that the domains for perfect isotropic growth and extremely anisotropic growth are near two regression lines  (see Figure~\ref{fig:MorphVar}f), \textit{i.e.}, $R_a = 0.045 T - 41.535$ and $R_a = 0.054 T - 56.442$, respectively.
That would be the key to the synthesis of large-area or extremely-anisotropic WS$_2$ monolayers.

\begin{figure*}[htb]
  \centering
  \includegraphics[width=0.75\textwidth]{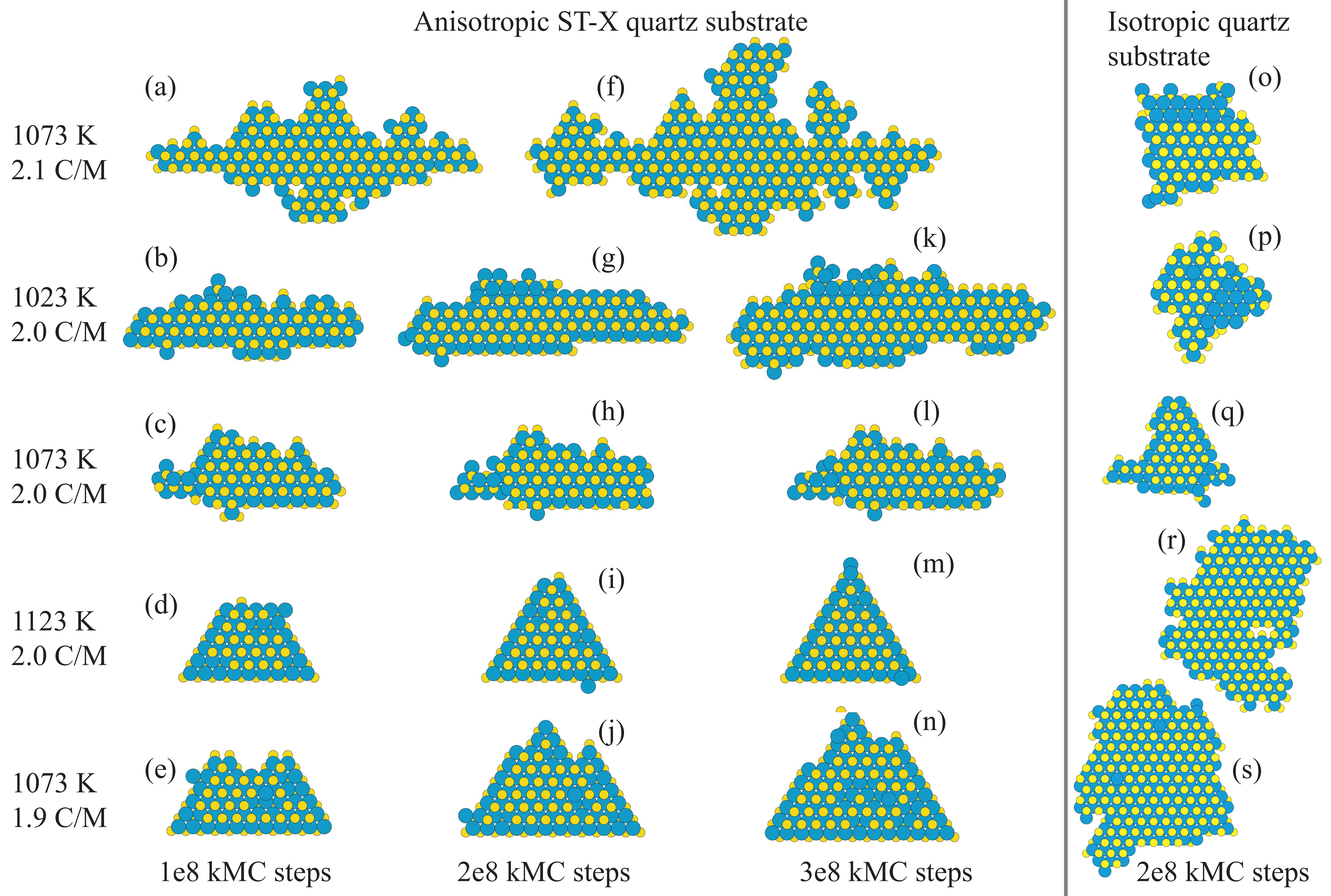}
  \caption{the simulated growth morphologies of monolayer WS$_2$ under diverse experiment conditions. (a-e) are the simulated results at 1e8 kMC steps, (f-j) at 2e8 kMC steps, (k-n) at 3e8 kMC steps, and (o-s) at 2e8 kMC steps; (a-n) are grown on anisotropic substrate while (o-s) on isotropic substrate. The gas flux is 9.0 W atoms/s.}
  \label{fig:FlakeTempCMRatio}
\end{figure*}

On the kMC perspective, there exists a competitive balance among the gas flux, C/M ratio, and temperature with regard to the impact on growth anisotropy. Figure~\ref{fig:MorphVar}a shows that the curve of T=1123 is similar to that of T=1173, which indicates that the temperature growth from 1123 K to 1173 K and the shift from 2.0 to 2.5 in the C/M ratio have the equivalent effect on growth anisotropy. Similarly, Figure~\ref{fig:MorphVar}f shows that the transition from isotropic to anisotropic zone can be achieved by increasing temperature or decreasing gas flux.
The evolution of growth morphology in essence is the movement of terminated edges.
And the type consistent of terminated edges determines whether the monolayer TMDC flakes can grow isotropically.
As shown in Figure~\ref{fig:FlakeTempCMRatio}m, the simulated flake is isotropically grown and it is terminated by 3 bottom-layer S-zz edges. In comparison, the simulated flake shown in Figure~\ref{fig:FlakeTempCMRatio}k has the extremely anisotropic growth morphology, which is outlined by 3 different types of terminated edges including bottom-layer S-zz, middle-layer W-zz, and top-layer S-zz.
At the vicinity of isotropic growth conditions, the AGR increase eventually as the terminated edges lose their type consistent to some extent.
That explains why the AGR does not increase monotonically as the C/M ratio grows up (see Figure~\ref{fig:MorphVar}a/c).

As for the controlled growth of nanoribbons of TMDC materials, we think it is feasible but still a challenge to obtain a reliable CVD product of TMDC nanoribbons though the preliminary experiment result presents a good and potential tendency to the controlled growth of elongated flakes.
At first, the two regression lines shown in Figure~\ref{fig:MorphVar}f are so close to each other that the morphological transition between anisotropy and isotropy is possible to occur frequently under the practical experimental conditions.
And secondly, the optimization of experimental parameters in the kMC simulation is relatively not limited while the CVD experiment is considerably limited by the feasibility and reliability of control of parameters.

Compared with the above-mentioned experiment results in Figure~\ref{fig:GrowthOfWS2}, the similar evolution is reproduced in the kMC simulation results.
Figure~\ref{fig:GrowthOfWS2} shows that flakes located at the upstream of gas flow have a higher AGR than those located at the downstream and the transition from trapezoid to triangle is observed on the sample at the downstream of gas flow.
Comparisons of the first, third, and fifth rows in Figure~\ref{fig:FlakeTempCMRatio} show that the growth morphology evolves from anisotropy to isotropy as the C/M ratio slightly decreases and explain why the transition from trapezoid to triangle is more likely to be observed at the downstream of gas flow.
Moreover, comparisons between the second, third, and fourth rows in Figure~\ref{fig:FlakeTempCMRatio} indicate that the increase of temperature also drives the morphological evolution from high anisotropy to almost perfect isotropy.
In our experimental installation, the concentration of W atoms is designed to be evenly distributed and that of S atoms would gradually decrease due to the consumption along the direction of gas flow.
The C/M ratio is thus likely to go down gradually along the direction of gas flow while the temperature variation at a small sample substrate can be ignored.~\cite{wang2014shape}
Consequently, the experiment results in Figure~\ref{fig:GrowthOfWS2} are much possibly attributed to the C/M ratio and, on the whole, the average AGR of monolayer WS$_2$ flakes is positively proportional to the C/M ratio.

The different effects of anisotropic and isotropic substrates on the growth of monolayer WS$_2$ also can be investigated by comparing the kMC simulation results.
Figure~\ref{fig:GrowthOfWS2}o-s show the simulated morphologies of monolayer WS$_2$ grown on the isotropic quartz substrate and the corresponding AGRs are about 1.0.
While, as shown in Figure~\ref{fig:GrowthOfWS2}a-n, the monolayer WS$_2$ flakes grown on the anisotropic ST-X quartz substrate tend to be expanded horizontally or along the X axis.
The comparisons suggest that the anisotropic growth of monolayer WS$_2$ is most possibly induced by the anisotropic property of the specific substrate.

In practice, AGR measurement curves and morphological classification maps provide a whole picture and detailed guide for operation.
The steep slope of AGR curves displayed in Figure~\ref{fig:MorphVar}a/c should be avoided to improve the controllability and reliability of experiments.
Because the gas flow is usually unidirectional during CVD process, the C/M ratio on the same substrate may fluctuate with the position.
The temperature and gas flux can be optimized according to the morphological classification maps shown in Figure~\ref{fig:MorphVar}b/d.
When the C/M ratio is significantly greater than 2.0, the varied range of C/M ratio is still away from the steep slope, therefore, 2D TMDC flakes of uniform morphology can be obtained.
In addition, although it is challenging to finely control the C/M ratio, the well control of the C/M ratio benefits much more than others.
Those two regression lines indicate that, when the C/M ratio is well controlled, the transition between anisotropy and isotropy can be easily made by adjusting either temperature or gas flux or both.
The above gives a whole picture of the dynamical morphologies in the anisotropic growth of monolayer WS$_2$ on ST-X quartz substrate.
Generally, those results and discussions can also apply to other TMDCs and even other 2D materials.

The anisotropic growth of monolayer TMDCs such as WS$_2$ is a critical step to reach the controlled growth, which would be a practical and effective approach to produce large single crystals and nanoribbons.
We studied the extremely anisotropic growth induced by the specific substrate using kMC simulations, proposed a quantitative model to measure the growth anisotropy and determine whether it is overgrown or undergrown, and provided a phenomenological picture of the morphological evolution of TMDC monolayers.
The simulation results not just predict the tendency of substrate-induced growth anisotropy but also provide the general technical requirments for the CVD equipment in order to produce TMDC nanoribbons.
For example, multi-zone temperature control and dual-direction gas flow control may be required to precisely control temperature and gas flow in the CVD furnace.

\section*{METHODS}
\subsection*{kMC simulations and DFT calculations}
\noindent The open-source codes, \textit{i.e.}, Atomic Simulation Environment (ASE)~\cite{larsen2017atomic} and Quantum ESPRESSO (QE)~\cite{QE-2009,QE-2017}, are utilized to prepare atomistic configurations and conduct density-functional theory (DFT) calculations.
The ASE suite was used to prepare the configuration of WS$_2$ and ST-X quartz.
The QE suite was used to conduct DFT calculations, where the bond energy is obtained by the regression of vacancy activation energies, the diffusion energy is calculated by the nudged elastic band method, and the adsorption energy is estimated by relaxation.

The kMC package, \textsf{kmos}~\cite{hoffmann2014kmos}, was used and the on-the-fly counting of nearest neighbors, including source atoms and substrate atoms, is enabled.
The kMC model has defined all physical processes involved in the CVD of monolayer TMDCs such as adsorption, desorption, and diffusion, and has also introduced local effects such as substrate adsorption and surface diffusion into those processes.

\subsection*{Growth and Characterization of WS$_2$ monolayers}
\noindent The substrate-induced anisotropic growth of WS$_2$ on ST-X quartz substrate (8~mm $\times$ 10~mm) was carried out in a quartz tube furnace (2 inch diameter).
The ST-X quartz substrate and SiO$_2$/Si wafers were cleaned in the sequence of acetone, isopropanol, and DI water for 10~min, respectively. 5~nm WO$_3$ films pre-deposited onto SiO$_2$/Si substrates via thermal evaporation of WO$_3$ powders (Alfa Aesar, 99.9\%) and S powders (Aladdin, 99.99\%) were used as precursors. Subsequently, WO$_3$-coated SiO$_2$/Si substrates were loaded into the tube together with the ST-X quartz substrates placed face-down above the former. A boat with 200~mg of S powder was placed outside the furnace and this zone was wrapped with a heating belt. After purging the quartz tube to $3.0 \times 10^{-2}$ Torr with a mechanical pump, the atmospheric pressure was recovered by introducing Ar gas. Then the substrate zone was heated to 875~\textdegree C at a rate of 20~\textdegree C/min while the S was heated to a temperature of 200~\textdegree C when the substrate zone reached 650~\textdegree C. The growth process was conducted at 875~\textdegree C for 30~min and then the furnace was cooled down to the room temperature naturally.

\subsection*{Data availability}
\noindent The data that support the findings of this study are available from the corresponding
author upon reasonable request.

\section*{ACKNOWLEDGEMENTS}
\noindent L. W. thanks J. M. Lorenzi for kindly help on the \textsf{kmos} kMC simulation package, and Y. Nie for fruitful discussion on the kMC simulation of the growth of monolayer TMDCs.
W. Y. and G. W. acknowledge support from the National Natural Science Foundation of China (Grants No. 61704040 and No. 61411136003).
L. W. acknowledges support from the program of China Scholarship Council (No. 201708330397).
  This research was also partially supported by Zhejiang Provincial Natural Science Foundation of China under Grant No. LGG19F040003.

\section*{ADDITIONAL INFORMATION}
\noindent\textbf{Supplementary information} accompanies the paper on the npj 2D Materials and Applications website.

\noindent\textbf{Competing interests:} The authors declare no competing interests.

\section*{AUTHOR CONTRIBUTIONS}
\noindent L. W. conceived the project and conducted the kMC simulations.
W. Y. performed the experiments (CVD growth and characterization of WS$_2$).
The manuscript was written through contributions of all authors.
All authors have given approval to the final version of the manuscript.

\section*{References}

\end{document}